\documentclass[12pt]{article}
\usepackage{latexsym}
\usepackage{amssymb}
\usepackage{amsthm}
\usepackage{epsfig}
\usepackage{graphicx}
\usepackage{mathtools} 
\usepackage{enumitem}
\usepackage{xcolor}
\usepackage{easyReview}
\usepackage{algorithm}
\usepackage[noend]{algpseudocode}

\usepackage{hyperref}
\hypersetup{
    linkcolor=blue,  
    linkbordercolor=red,
    linktoc=all,     
    linktocpage
}

\def\dsp{\def\baselinestretch{1.5}\large\normalsize}
\dsp
\allowdisplaybreaks
\begin{document}
\title{Asymptotic Solution for Skin Heating by an Electromagnetic Beam at an Incident Angle}
\author{Hongyun Wang\footnote{Corresponding author, hongwang@ucsc.edu} 
\\
Department of Applied Mathematics \\
University of California, Santa Cruz, CA 95064, USA\\
\\
Shannon E. Foley \\
U.S. Department of Defense \\
Joint Intermediate Force Capabilities Office \\
Quantico, VA 22134, USA\\
\\
Hong Zhou \\ Department of Applied Mathematics\\
Naval Postgraduate School, Monterey, CA 93943, USA
}

\maketitle

\begin{abstract}
We investigate the temperature evolution in the three-dimensional skin tissue exposed to a millimeter-wave electromagnetic beam that is not necessarily perpendicular to the skin surface. 
This study examines the effect of the beam's incident angle. 
The incident angle influences the thermal heating in two aspects: 
(i) the beam spot projected onto the skin is elongated compared to the intrinsic beam spot in 
a perpendicular cross section, resulting in a lower power per skin area; and
(ii) within the tissue, the beam propagates at the refracted angle relative to the depth direction. At millimeter-wavelength frequencies, the characteristic penetration depth is sub-millimeter, 
whereas the lateral extent of the beam spans at least several centimeters in applications. 
We explore the small ratio of the penetration depth to the lateral length scale 
in a non-dimensional formulation and 
derive a leading-term asymptotic solution for the temperature distribution. 
This analysis does not rely on a small incident angle and is therefore applicable to 
arbitrary angles of incidence.
Based on the asymptotic solution, we establish scaling laws for the three-dimensional skin 
temperature, the skin surface temperature, and the skin volume in which thermal nociceptors 
are activated. 
\end{abstract}


\clearpage
\renewcommand*\contentsname{Table of contents}
\tableofcontents
\clearpage

\section{Introduction}
Millimeter-wave (MMW) systems operate with electromagnetic beams
 in the frequency range of 30-300 GHz. 
These systems play a significant role in both defense and civilian applications. The biological effects of MMW exposure -- particularly the thermal effects on human skin -- have been
studied extensively. A number of these studies were compiled in a special issue of the Journal of Directed Energy \cite{Whitmore_2021, Miller_2021, Cook_2021, Cook_2021B, Haeuser_2021, Cobb_2021, Parker_2021, Parker_2021B}. 

A primary effect of MMW exposures is skin temperature increase heated by the absorbed electromagnetic power in the tissue. 
At frequencies between 30 and 300 GHz, the electromagnetic power is almost entirely 
absorbed within less than one millimeter of skin depth. This results in a rapid temperature 
rise in a thin skin layer and creates a steep temperature gradient in the depth direction. 
In the lateral directions along skin surface, the heating varies with the power density 
across the beam cross section, which typically spans several centimeters on the skin surface. 
This pronounced difference in spatial scales between the depth and lateral directions 
renders lateral heat conduction negligible compared to conduction in the depth direction.
When the incident electromagnetic beam is perpendicular to the skin surface, 
the separation of length scales allows for a simplified model of skin temperature: 
a one-dimensional partial differential equation governing heat conduction in the depth direction. 
Many previous theoretical studies have focused on this perpendicular-beam case. 
The current study examines the thermal effects induced by an electromagnetic beam 
at an arbitrary incident angle. 

This paper is organized as follows. 
Section 2 describes beam propagation both outside and within the skin, the absorption of electromagnetic power by the tissue, and the governing equation for the evolution of skin temperature.
In Section 3, we derive an asymptotic solution to the governing equation.
The analysis does not assume a small incident angle; rather, 
it is based on that the depth scale is much smaller than the length scale in lateral directions. 
Consequently, the asymptotic solution is valid for arbitrary incident angles. 
Section 4 explores the scaling behavior of the three-dimensional skin temperature, the skin surface temperature, and the activated skin volume. These scaling laws provide a foundation for comparing the effects of beams with different beam configurations. Finally, Section 5 summarizes the main results of the study. 

%
\section{Governing Equation for Skin Temperature} \label{temp_evolu}
\subsection{Coordinate Systems and the Incident Beam}
\begin{figure}[!h]
\vskip 0.4cm
\begin{center}
\psfig{figure=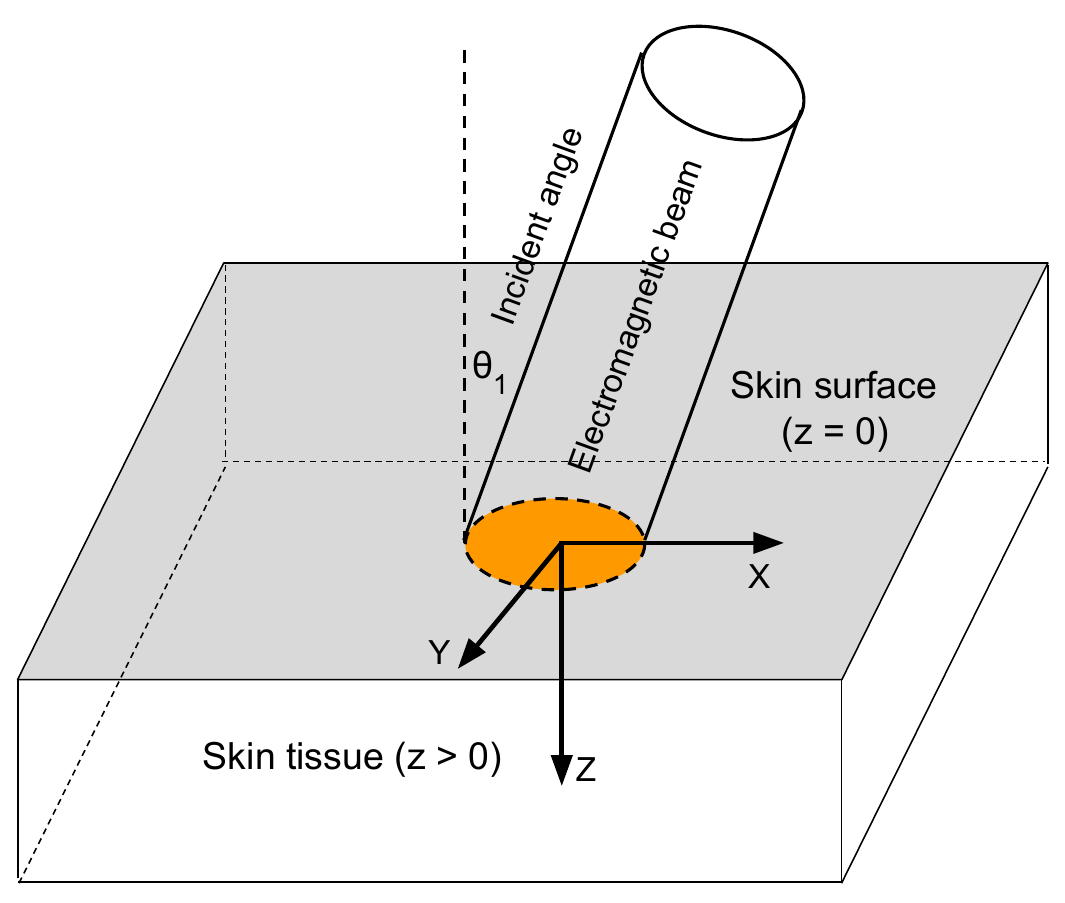, height=3.5in} 
\end{center}
\vskip -0.8cm
\caption{Schematic diagram of the coordinate system and
the incident beam. } 
\label{fig_01}
\end{figure}

We consider the case of a flat skin surface and establish a coordinate system similar 
to that used in our previous study \cite{Wang_2020}. As illustrated in Figure \ref{fig_01}, 
the normal direction into the skin tissue is defined as the positive $z$-axis, with 
$z=0$ at the skin surface. The $z$-coordinate represents the depth into the skin.
The incident beam is tilted away from the $z$-axis by an angle $\theta_1$, 
referred to as the incident angle. This tilt occurs within the $(x, z)$-plane, with the positive $x$-axis indicating the direction of the tilt. The positive $y$-axis is defined accordingly, completing the right-handed coordinate system. 

We consider the situation where the power density over a 
perpendicular cross-section of the beam follows a general 2D Gaussian distribution. 
The general 2D Gaussian allows 
arbitrary aspect ratios in the perpendicular cross-section. 
Note that when the incident beam is tilted, the intersection 
of the beam with the skin surface is an angled cross-section, not
a perpendicular one. 

Given the 2D Gaussian power distribution, the incident beam is 
characterized by its configuration, which includes both the tilt of the  
beam axis in the $(x, z)$-plane and the rotation of the elliptical 2D Gaussian distribution 
about the beam axis. 
Let $\phi$ be the azimuthal angle about the beam axis, between the $(x, z)$-plane 
and the major axis of the elliptical 2D Gaussian distribution over a perpendicular cross-section.  
We introduce three beam configurations:
\begin{itemize}
\item C-0 is the simple configuration in which the beam axis is aligned with 
the $z$-axis (i.e., zero incident angle), and the major axis of 
the elliptical 2D Gaussian distribution is aligned with the $x$-axis 
(i.e., zero azimuthal angle). 
\item C-1 is the configuration obtained by rotating C-0 about the positive 
 $z$-axis (the beam axis in C-0, going into the skin) \underline{clockwise} 
 by the azimuthal angle $\phi $. 
\item C-2 is the configuration obtained by rotating C-1 about the 
positive $y$-axis \underline{clockwise} by the incident angle $\theta_1$ 
(tilting the beam away from the $z$-direction). 
\end{itemize}
C-2 is the beam configuration shown in Figure \ref{fig_01}, although the azimuthal 
angle $\phi$ is not explicitly illustrated.  The main objective of this paper is to study
the thermal effect of a beam in configuration C-2.
As described above, C-2 is obtained from C-0 through two successive rotations. 

\subsection{Power Density Projected onto the Skin Surface}
\begin{figure}[!h]
\vskip 0.6cm
\begin{center}
\psfig{figure=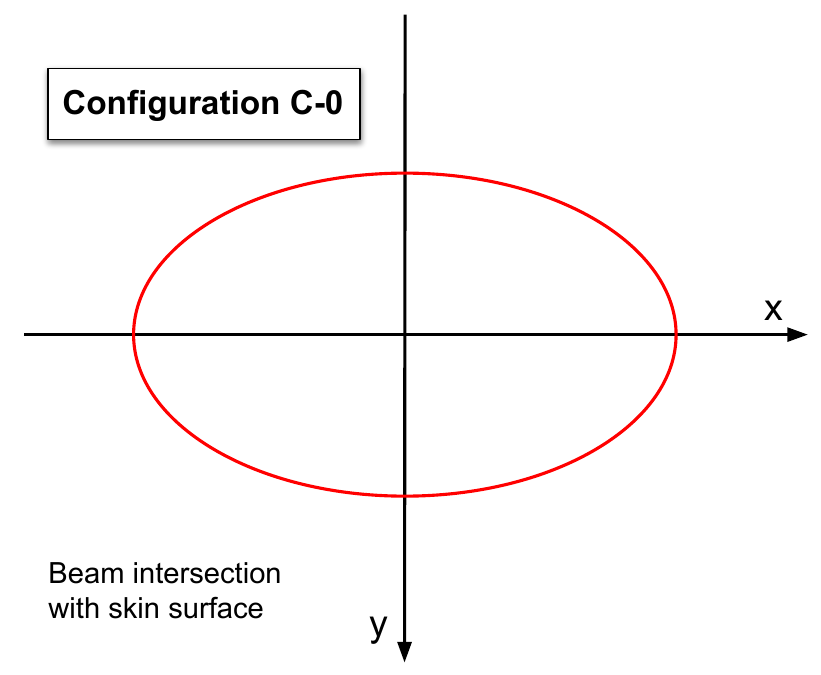, width=2.8in} \;\;\;\;
\psfig{figure=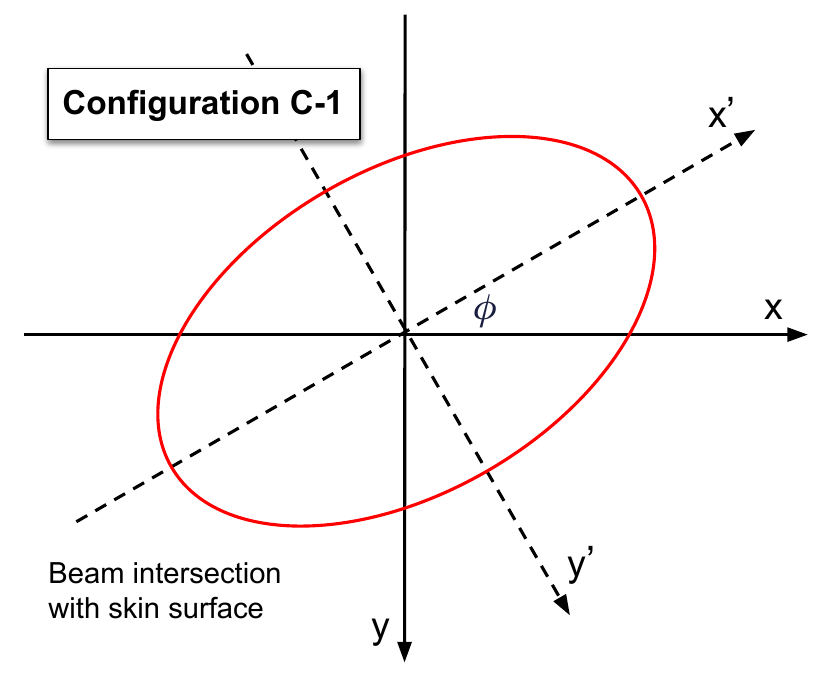, width=2.8in} \\[2ex]
\psfig{figure=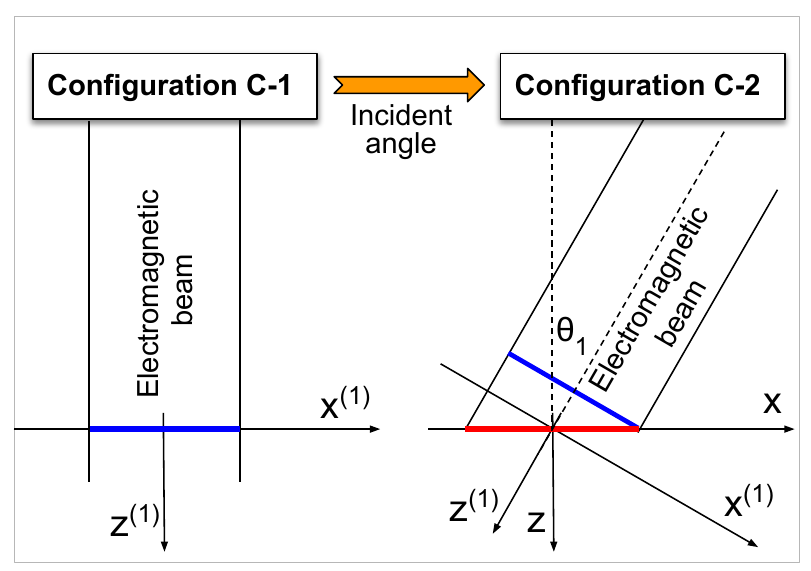, width=3.2in} \;\;
\psfig{figure=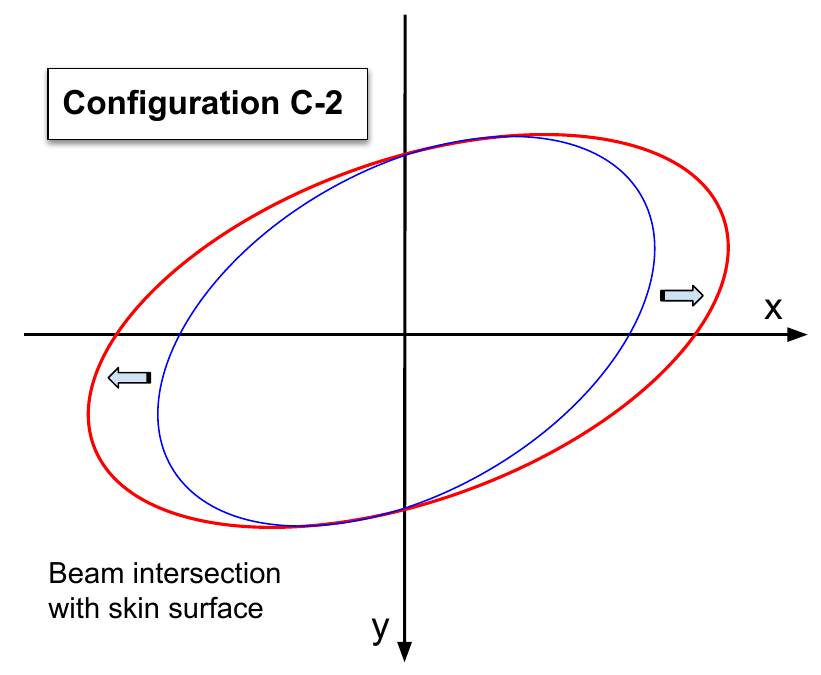, width=2.8in}
\end{center}
\vskip -0.6cm
\caption{Transition from beam configuration C-0 to C-2. 
Top left: The beam intersection with skin surface in C-0, represented by its 
$e^{-1/2}$-maximum contour of the power density. 
Top right: The intersection of C-1. 
Bottom left: The tilting of beam from C-1 to C-2.
Bottom right: The intersection of C-2.} 
\label{fig_02}
\end{figure}

In each of the three beam configurations introduced above, 
we examine the intersection of the beam with the skin surface
and the power density over this intersection as a function of $(x, y)$. 
Here the power density refers to the power per unit area of the intersection with 
the skin surface, which may not necessarily be a perpendicular cross-section of the beam. 
In configuration C-0, the intersection with the skin surface is a perpendicular cross-section,
and the power density is given by:
\begin{equation}
P_d^\text{(C-0)}(x, y) = P_d^{(i)} \exp\Big(\frac{-1}{2}(\frac{x^2}{\sigma_1^2}+\frac{y^2}{\sigma_2^2})\Big)
\label{Pd_0}
\end{equation}
where $P_d^{(i)}$ is the power density at the beam center over a perpendicular 
cross-section. We refer to $P_d^{(i)}$ as the intrinsic beam power density, 
as it is independent of the incident angle. 
Over the beam intersection with the skin surface, we represent 
the relative distribution of the power density concisely by using its contour 
at the level of $(e^{-1/2}{\times} \text{maximum})$. 
For the beam in configuration C-0, the $e^{-1/2}$-maximum contour is 
shown in the top-left panel of Figure \ref{fig_02}. This contour is the ellipse described by
\( \frac{x^2}{\sigma_1^2}+\frac{y^2}{\sigma_2^2} = 1 \), 
where $\sigma_1$ and $\sigma_2$ are respectively, the semi-major and semi-minor axes, 
of the perpendicular beam cross-section.

In configuration C-1, the beam intersection with the skin surface is the same perpendicular 
cross-section as in configuration C-0 except that it is rotated about the beam axis 
by an angle $\phi$, as shown in the top-right panel of Figure \ref{fig_02}. 
The rotation is clockwise about the positive $z$-axis (into the page). 
The power density and the $e^{-1/2}$-maximum contour remain unchanged 
in the local coordinate system $(x', y')$ attached to the beam. 
The global coordinates $(x, y)$ and the local coordinates $(x', y')$ are related by the following transformation:
\begin{equation}
\begin{dcases}
x' = x \cos\phi - y\sin\phi \\
y' = x \sin\phi + y\cos\phi 
\end{dcases}
\label{x_to_x'}
\end{equation}
In coordinates $(x, y)$, the power density of configuration C-1 is 
\begin{align}
& P_d^\text{(C-1)}(x, y) = P_d^\text{(C-0)}(x', y') \nonumber \\
& \qquad = P_d^{(i)} \exp\Big(\frac{-1}{2}(\frac{\big(x \cos\phi - y\sin\phi \big)^2}{\sigma_1^2}
+\frac{\big(x \sin\phi + y\cos\phi \big)^2}{\sigma_2^2})\Big) 
\label{Pd_1B}
\end{align}

The rotation from configuration C-1 to C-2 is illustrated in 
the bottom-left panel of Figure \ref{fig_02}. 
The beam intersection with the skin surface in C-1 (solid blue line) is 
a perpendicular cross-section, while the intersection in C-2
(solid red line) is an angled cross-section at angle $\theta_1$ 
relative to the perpendicular cross-section. 
These two intersections are connected by beam ray lines (lines parallel to the beam axis). 
A point $(x, y)$ in the intersection of C-2 is traced along a beam ray line to the
point $(x \cos\theta_1, y)$ in the intersection of C-1. 
An area element at point $(x \cos\theta_1, y)$ in the intersection of C-1
is mapped by beam ray lines to an area element at point $(x, y)$ in the intersection of C-2. 
This mapping also magnifies the area element by a factor of $(1/\cos\theta_1)$ as it moves 
from C-1 to C-2. 

Over the intersection with the skin surface, we represent the relative distribution 
of power density using its $e^{-1/2}$-maximum contour in the $(x, y)$-plane. 
We refer to this $e^{-1/2}$-maximum contour simply as the beam spot, or more precisely, 
the beam spot projected onto the skin surface. 
The beam spot of C-2 (red ellipse) and the beam spot of C-1 (blue ellipse) 
are compared in the bottom-right panel of Figure \ref{fig_02}. 
Essentially, the beam spot of C-2 is obtained by stretching the beam spot of C-1
in the $x$-direction by a factor of $(1/\cos\theta_1)$.
When the beam is in configuration C-2, the power density over the intersection
with the skin surface can be written as
\begin{align}
& P_d^\text{(C-2)}(x, y) = \cos\theta_1 P_d^\text{(C-1)}(x \cos\theta_1, y) 
\nonumber \\
& \qquad =  (\cos\theta_1 P_d^{(i)}) \exp\Big(\frac{-1}{2}(\frac{\big(x \cos\theta_1\cos\phi - y\sin\phi \big)^2}{\sigma_1^2}
+\frac{\big(x \cos\theta_1 \sin\phi + y\cos\phi \big)^2}{\sigma_2^2})\Big) 
\nonumber \\
& \qquad \equiv (\cos\theta_1  P_d^{(i)}) \exp\Big( \frac{-1}{2}
\big( a_{11} x^2 + a_{22} y^2 + 2 a_{12} x y \big) \Big) 
\label{Pd_2A}
\end{align}
where the coefficients $a_{11}$, $a_{22}$, and $a_{12}$ are functions of 
$(\sigma_1, \sigma_2, \phi, \theta_1)$. 
\begin{equation}
\begin{dcases}
a_{11}(\sigma_1, \sigma_2, \phi, \theta_1) = \cos^2\theta_1\Big(\frac{\cos^2\phi}{\sigma_1^2}
+\frac{\sin^2\phi}{\sigma_2^2}\Big) \\
a_{22}(\sigma_1, \sigma_2, \phi, \theta_1) = \frac{\sin^2\phi}{\sigma_1^2}+\frac{\cos^2\phi}{\sigma_2^2} \\
a_{12}(\sigma_1, \sigma_2, \phi, \theta_1) = \cos\theta_1 \sin\phi \cos\phi\big(\frac{1}{\sigma_2^2}-\frac{1}{\sigma_1^2}\big) 
\end{dcases} 
\label{a2_coefs}
\end{equation}
\eqref{Pd_2A} is the general form of the 2D Gaussian distribution. 
Let $(\xi, \eta)$ be the rotated coordinate system in which the
Gaussian distribution \eqref{Pd_2A} is diagonalized. We then have 
\begin{align}
& \begin{dcases}
\xi = x \cos\phi_2 - y \sin\phi_2 \\
\eta = x \sin\phi_2 + y \cos\phi_2 
\end{dcases} 
\label{xi_eta} \\[1ex]
& a_{11} x^2 + a_{22} y^2 + 2 a_{12} x y  
= \frac{\xi^2}{\sigma_{\xi}^2}+\frac{\eta^2}{\sigma_{\eta}^2} 
\label{Pd_2B}
\end{align}
where $\phi_2$ is the azimuthal orientation angle of the beam spot projected onto 
the skin surface. $\phi_2$ is the clockwise angle about the positive $z$-axis
(into the skin) from the $x$-axis to the major axis. 
$(\sigma_{\xi}, \sigma_{\eta})$ are the half-major and half-minor axes
of the beam spot projected onto the skin surface.  
The projected beam spot is completely described by geometric parameters 
$(\sigma_{\xi}, \sigma_{\eta}, \phi_2)$.
We express $(\sigma_{\xi}, \sigma_{\eta}, \phi_2)$ in terms of 
$(\sigma_{\xi}, \sigma_{\eta}, \phi_2)$. 
Substituting \eqref{xi_eta} into \eqref{Pd_2B} to express both sides as 
quadratic forms of $(x, y)$ and equating the corresponding coefficients on both sides, 
we derive a set of equations for $(\sigma_{\xi}, \sigma_{\eta}, \phi_2)$.
\begin{equation}
\begin{dcases}
 \frac{\cos^2\phi_2}{\sigma_{\xi}^2}+
 \frac{\sin^2\phi_2}{\sigma_{\eta}^2} = a_{11} \\
 \frac{\sin^2\phi_2}{\sigma_{\xi}^2}+
 \frac{\cos^2\phi_2}{\sigma_{\eta}^2} = a_{22} \\
\sin\phi_2 \cos\phi_2
\big(\frac{1}{\sigma_{\eta}^2}-\frac{1}{\sigma_{\xi}^2}\big)  = a_{12}
\end{dcases} 
\label{sys_q_s1_s2}
\end{equation}
We solve \eqref{sys_q_s1_s2} for $(\phi_2, \sigma_{\xi}, \sigma_{\eta})$ 
in terms of the coefficients $(a_{11}, a_{22}, a_{12})$ and arrive at:
\begin{equation}
\begin{dcases} 
\phi_2(\sigma_1, \sigma_2, \phi, \theta_1) = \text{atan2}\Big(2a_{12}, \; (a_{22}-a_{11}) +
\sqrt{(a_{22}-a_{11})^2+(2a_{12})^2}\Big) \\[1ex]
\sigma_{\xi}(\sigma_1, \sigma_2, \phi, \theta_1) = \frac{1}{\sqrt{2 (a_{11}a_{22}-a_{12}^2)} }  
\sqrt{ (a_{22}+a_{11}) + \sqrt{(a_{22}-a_{11})^2+(2a_{12})^2}} \\[1ex]
\sigma_{\eta}(\sigma_1, \sigma_2, \phi, \theta_1) = \frac{1}{\sqrt{2 (a_{11}a_{22}-a_{12}^2)} }    
\sqrt{ (a_{22}+a_{11}) - \sqrt{(a_{22}-a_{11})^2+(2a_{12})^2}}
\end{dcases}
\label{phi2_s1_s2}
\end{equation} 
where $(a_{11}, a_{22}, a_{12})$ are calculated from 
$(\sigma_1, \sigma_2, \phi, \theta_1)$ in \eqref{a2_coefs}.
Equation \eqref{phi2_s1_s2} defines the mapping 
from the given parameters of beam intrinsic geometry and beam orientation
$(\sigma_1, \sigma_2, \phi, \theta_1)$ to the geometric parameters 
of the beam spot projected onto the skin surface $(\sigma_{\xi}, \sigma_{\eta}, \phi_2)$. 
This projection mapping alters both the aspect ratio and the azimuthal
orientation angle of the elliptic beam spot in a non-trivial manner. 
However, the areas of the beam spot before and after the projection are related in a simple way, namely:
\begin{equation}
\sigma_\xi \sigma_\eta = \frac{1}{\cos\theta_1} \sigma_1 \sigma_2
\label{area_proj}
\end{equation}

In an exposure event, the beam setup (which includes the intrinsic beam itself and 
its orientation relative to the skin) is completely described by five parameters 
$(P_d^{(i)}, \sigma_1, \sigma_2, \phi, \theta_1)$, where: 
i) $\theta_1$ is the incident angle; 
ii) $P_d^{(i)}$ is the beam center power density over a perpendicular 
cross-section;  
iii) $(\sigma_1, \sigma_2)$ are the half-major and half-minor axes of the power distribution 
over a perpendicular cross-section; and
iv) $\phi$ is the azimuthal orientation angle about the beam axis
between the major axis and the $(x, z)$ plane. 
The parameters $(P_d^{(i)}, \sigma_1, \sigma_2, \phi)$ are intrinsic beam parameters, 
independent of the incident angle. 
For the beam setup $(P_d^{(i)}, \sigma_1, \sigma_2, \phi, \theta_1)$, 
the power density projected onto the skin surface is 
\begin{equation}
\boxed{\qquad 
\begin{dcases}
 P_d^\text{(proj)}(x, y) = P_d^\text{(proj)}{\cdot} f(x, y), \\[1ex]
\qquad P_d^\text{(proj)} \equiv (\cos\theta_1  P_d^{(i)}), \\
\qquad f(x, y; \sigma_{\xi}, \sigma_{\eta}, \phi_2) \equiv \exp\Big( \frac{-1}{2}
\big( \frac{\xi^2}{\sigma_{\xi}^2}+\frac{\eta^2}{\sigma_{\eta}^2}\big) \Big), \\
\qquad \xi = x \cos\phi_2 - y \sin\phi_2, \quad \eta = x \sin\phi_2 + y \cos\phi_2 
 \end{dcases} \qquad }
 \label{Pd_eff}
\end{equation}
where $P_d^\text{(proj)} \equiv (\cos\theta_1  P_d^{(i)})$ is the 
beam center power density projected onto the skin surface, and 
$f(x, y)$ is the relative distribution of the power density over the skin surface. 
$f(x, y)$ is described by the geometric parameters $(\sigma_\xi, \sigma_\eta, \phi_2)$. 
In this study, we reserve the notation $(\sigma_\xi, \sigma_\eta, \phi_2)$ 
for the values calculated from $(\sigma_1, \sigma_2, \phi, \theta_1)$
using \eqref{phi2_s1_s2} and \eqref{a2_coefs}. 

We consider the following two beam setups.
\begin{equation}
\begin{dcases}
\text{The original beam setup at a nontrivial incident angle $\theta_1$:} \\
\hspace*{1cm} (P_d^{(i)}, \; \sigma_1, \; \sigma_2, \; \phi, \; \theta_1) \\[1ex]
\text{The ``projected'' beam setup that is perpendicular to skin surface:} \\
\hspace*{1cm} 
\big( P_d^\text{(proj)}, \; \sigma_\xi, \; \sigma_\eta, \; \phi_2, \; \theta_1= 0 \big) \\
\hspace*{1cm}  P_d^\text{(proj)} = (\cos\theta_1 P_d^{(i)}), \qquad 
 \big[\sigma_\xi, \sigma_\eta, \phi_2 \big](\sigma_1, \sigma_2, \phi, \theta_1) 
 \text{ given in \eqref{phi2_s1_s2}.}
\end{dcases} \label{projected_setup}
\end{equation}
\begin{figure}[!h]
\vskip 0.6cm
\begin{center}
\psfig{figure=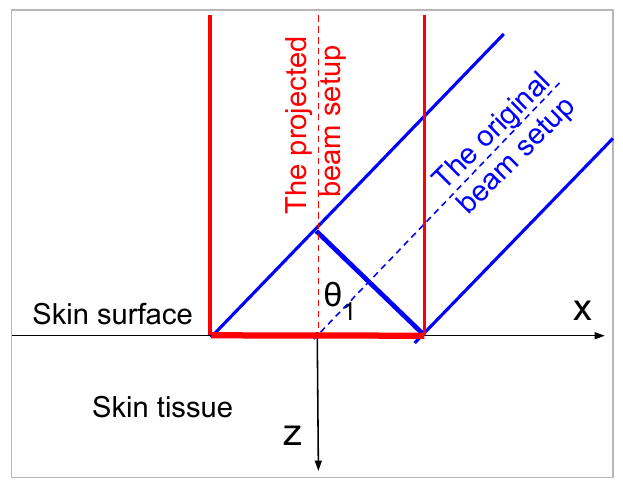, width=3.5in}
\end{center}
\vskip -0.6cm
\caption{The original beam setup and the projected beam setup, as defined in 
\eqref{projected_setup}.} 
\label{fig_02B}
\end{figure}
The two beam setups are illustrated in Figure \ref{fig_02B}. The original beam setup and the projected beam setup have the same projected power density on the skin surface. The original setup is at an incident angle, while the projected setup is perpendicular to the skin surface. However, they are not equivalent. The effect of the incident angle extends beyond merely influencing the power density projected onto the skin surface. In Snell's law, the incident angle determines the angle of beam propagation inside the skin tissue, which in turn affects the power absorption 
and the corresponding heating source in the skin.
\subsection{Beam Propagation and Power Absorption in Skin}
As given in \eqref{Pd_eff}, for a Gaussian beam at an incident angle $\theta_1$, 
the power density projected onto the skin surface retains a two-dimensional Gaussian profile, albeit
with a reduced beam center power density and elongated beam spot  
due to the oblique incidence. 
We examine the propagation and absorption of electromagnetic power as the beam 
penetrates the skin surface and continues into the underlying tissue layers.

\begin{figure}[!h]
\vskip 0.6cm
\begin{center}
\psfig{figure=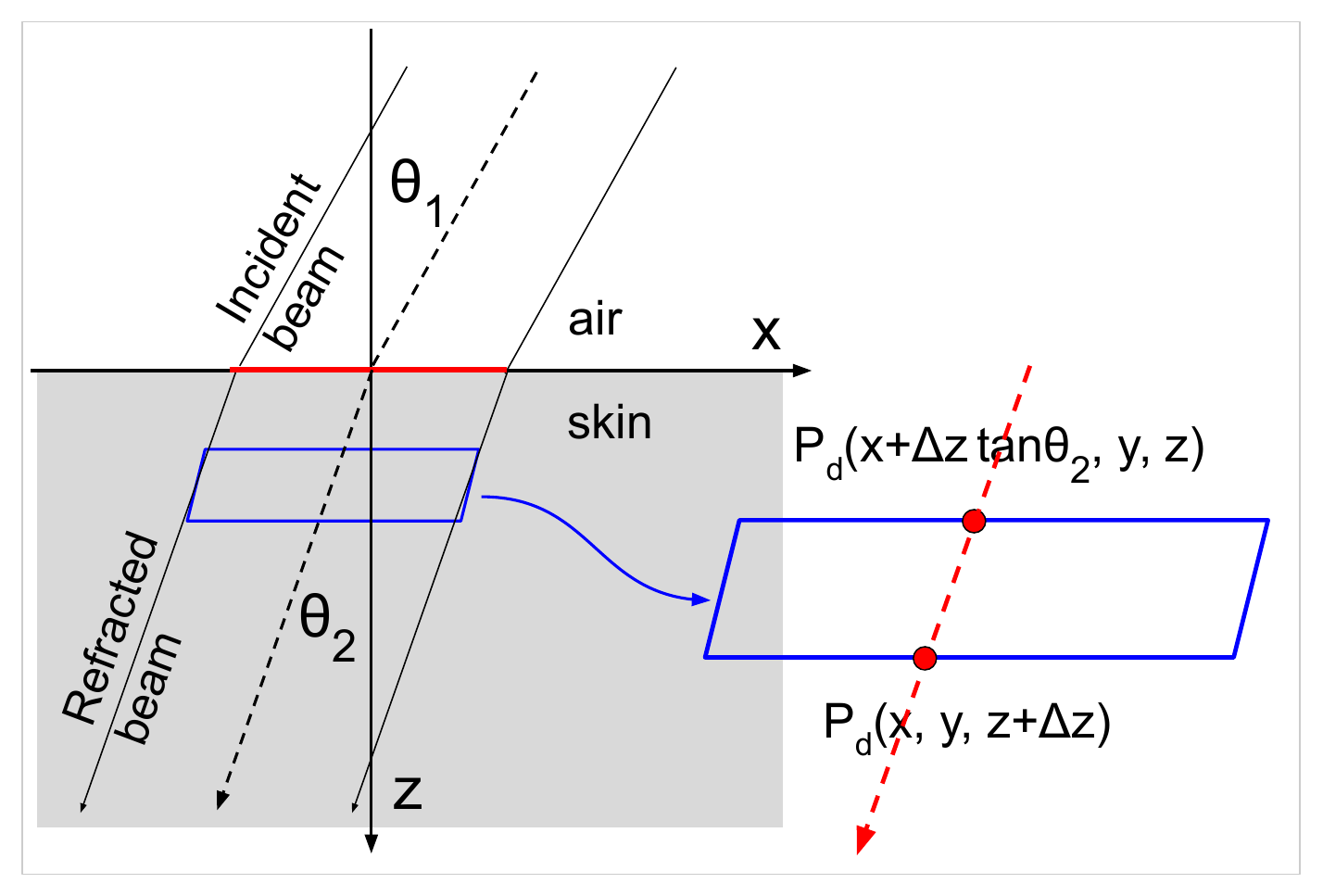, width=4.0in} 
\end{center}
\vskip -0.6cm
\caption{Beam propagation and absorption inside skin after refraction. 
$P_d(x, y, z)$ is the power per unit area passing through the plane 
that is parallel to the skin surface at depth $z$.} 
\label{fig_03}
\end{figure}

Let $\alpha $ be the fraction of the incident beam power that passes through the skin surface
(the remaining power is reflected). 
The power density passing through the skin surface is therefore given by 
$\alpha P_d^\text{(proj)}(x, y)$ \cite{Cazares_2019}. 
Let $\theta_2 $ denote the angle of refraction, defined as the angle between the $z$-axis 
and the direction of beam propagation inside the skin (see Figure \ref{fig_03}). 
The incident angle $\theta_1 $ and the refracted angle $\theta_2 $ are related by Snell's law:
\begin{equation}
\frac{\sin\theta_1}{\sin\theta_2} = \frac{n_\text{skin}}{n_\text{air}}
\label{Snell_law}
\end{equation}
where $n_\text{skin} \approx 1.4$ and $n_\text{air} \approx 1.0$ are the refractive indices of skin and air, respectively \cite{Anderson_1981, Van_Gemert_1989, 
Bezugla_2024}. Since $n_\text{skin} > n_\text{air}$, the refracted angle 
is smaller than the incident angle, i.e., $\theta_2 < \theta_1$.

Let $P_d(x, y, z)$ denote the power flux at position $(x, y, z)$, defined as the power per area
passing through the plane parallel to the skin surface at depth $z$.
Note that this plane is not perpendicular to the beam axis. Thus,
$P_d(x, y, z)$ represents the power per unit area on an angled cross-section 
that is obliquely oriented relative to the beam's direction of propagation (see Figure \ref{fig_03}).
Inside the skin, the beam rays follow straight-line paths described by 
\( x(z) = x_\text{surf}-z \tan\theta_2, \;\; y(z) = y_\text{surf}\). 
For the beam ray passing through point $(x, y, z)$ inside the skin, the corresponding 
entry point on the skin surface can be determined by tracing the beam ray backward
to its intersection with the skin surface. This gives the following relations:
\[ x_\text{surf} = x+z \tan\theta_2,\quad y_\text{surf} = y \] 
At the skin surface ($z=0$), using $P_d(x, y, 0) = \alpha P_d^\text{(proj)}(x, y)$
and \eqref{Pd_eff}, we have  
\begin{equation}
\begin{aligned}
& P_d(x, y, 0) =  P_d^{(a)} {\cdot} f(x, y, \sigma_\xi, \sigma_\eta, \phi_2) \\ 
& \qquad P_d^{(a)} \equiv (\alpha P_d^\text{(proj)}) = (\alpha \cos\theta_1  P_d^{(i)}) 
\end{aligned}
\label{Pd_absorb}
\end{equation}
Here, $P_d^\text{(proj)}$ represents the beam center power density arriving at the skin surface, while
$P_d^{(a)}$ is the beam center power density that passes through the skin surface 
and is subsequently absorbed into the skin.
$P_d^{(i)}$ denotes the beam center power density through a perpendicular 
cross-section in air, which is an intrinsic beam property independent of the incident angle. It is important to note that the skin surface is perpendicular neither to 
the incident beam in air nor to the refracted beam inside the skin. 
$f(x, y, \sigma_\xi, \sigma_\eta, \phi_2)$ is the relative distribution 
of the power density projected onto the skin surface given in \eqref{Pd_eff}. 

As the beam propagates through the skin, some electromagnetic power is absorbed. At depth $z$, $P_d(x, y, z) $ represents the remaining portion of $P_d(x+z \tan\theta_2,  y, 0) $ from $z=0$ that has survived the absorption. According to the
Beer-Lambert Law, the power absorption per unit length of propagation
is proportional to the remaining power. 
From the skin surface to depth $z$, the propagation distance of 
the refracted beam is $(z/\cos\theta_2) $. Appying the
Beer-Lambert Law, the power density at depth $z$ is given by:
\begin{equation}
P_d(x, y, z)  = 
P_d(x+z \tan\theta_2,  y, 0) e^{-\frac{\mu}{\cos\theta_2} z}
= P_d^{(a)} {\cdot} f(x+z \tan\theta_2, y) e^{-\frac{\mu}{\cos\theta_2} z}
\label{Pd_z_decay}
\end{equation}
where $\mu $ is the skin absorption coefficient for the electromagnetic 
frequency used. 
The propagation of the refracted beam and the power density $P_d(x, y, z)$ inside the skin are 
illustrated in Figure \ref{fig_03}. 
We calculate the power absorbed per skin volume, which becomes the heating source. 
Consider a small parallelepiped defined by three pairs of parallel planes:
\( y=y_c\pm(\Delta y)/2\), \(z=z_c\pm(\Delta z)/2\), and 
\(x=x_c\pm(\Delta x)/2+z \tan\theta_2 \). 
The volume of this parallelepiped is $(\Delta x)(\Delta y) (\Delta z)$. 
The net power going into the parallelepiped is obtained by 
differentiating the power flux $P_d(x, y, z)$ with respect to $z$. 
\begin{align*}
& (\Delta x)(\Delta y) P_d(x_c, y_c, z)\Big|_{z=z_c+(\Delta z)/2}^{z=z_c-(\Delta z)/2}
\approx -(\Delta x)(\Delta y) (\Delta z) \frac{\partial P_d(x_c, y_c, z)}{\partial z}
 \Big|_{z=z_c} \\
& \qquad \approx (\Delta x)(\Delta y) (\Delta z) 
P_d^{(a)} {\cdot} f(x_c+z_c \tan\theta_2, y_c)  \frac{\mu}{\cos\theta_2}
e^{-\frac{\mu}{\cos\theta_2}z}
\end{align*}
Based on conservation of energy, the power absorbed per volume at $(x, y, z)$ is 
given by the net power going into the small parallelepiped divided by its volume
$(\Delta x)(\Delta y) (\Delta z) $: 
\begin{equation}
\boxed{\quad \big(\begin{array}{c}
\text{power absorbed} \\[-2ex]
\text{per volume} \end{array} \big)
=P_d^{(a)} {\cdot} f(x+z \tan\theta_2, y) \frac{\mu}{\cos\theta_2} 
e^{-\frac{\mu}{\cos\theta_2} z} \quad}
\label{Pd_vol}
\end{equation}
\eqref{Pd_vol} gives the heat source in the skin temperature evolution. Note that 
\eqref{Pd_vol} decays exponentially with respect to $z$, with a characteristic penetration 
depth of $1/\mu$. 
Therefore, the effect of the heat source is practically confined to a few multiples of $1/\mu$
in depth. 

\subsection{Skin Temperature Evolution}
Let $T(x, y, z, t)$ denote the skin temperature at position $(x, y, z)$ at time $t$. 
To establish the governing equation for $T(x, y, z, t)$, 
we proceed with the following assumptions. 
\begin{enumerate}
    \item The skin's material properties are uniform in space, independent of $(x, y, z)$.
    \item The skin temperature prior to electromagnetic exposure is uniform in space, which  
    is called the baseline skin temperature and denoted by $T_\text{base}$. 
    \item The electromagnetic power input 
per area at the skin surface is much larger than the rate of heat loss at the skin surface
due to black body radiation, evaporation or convective cooling by air. 
As a result, in the model formulation, we neglect the heat loss at the skin surface. 
     \item The thickness of skin tissue is much larger than both the electromagnetic penetration depth and the thermal diffusion depth over the short exposure duration. As a result, 
     in the model formulation, we treat the skin as a semi-infinite domain 
     extending in the depth direction to $z = +\infty$. 
\end{enumerate}
The governing equation for $T(x, y, z, t)$ is derived from conservation of energy. 
The initial and boundary conditions for $T(x, y, z, t)$ follow directly from Assumptions 2 and 3.
\begin{equation}
\begin{dcases}
\underbrace{\rho_m C_p\frac{\partial T}{\partial t}}_{\substack{
\text{rate of change}\\ \text{of heat in skin}}}
= \underbrace{k \Big(\frac{\partial^2T}{\partial x^2}
+\frac{\partial^2T}{\partial y^2}+\frac{\partial^2T}{\partial z^2}
\Big)}_{\substack{
\text{heat flux from}\\ \text{conduction}}} 
+ \underbrace{P_d^{(a)} {\cdot} f(x+z \tan\theta_2, y) \frac{\mu}{\cos\theta_2} 
e^{-\frac{\mu}{\cos\theta_2} z}}_{
\substack{\text{heat source from}\\ \text{absorbed power}}} \\[1ex]
\frac{\partial T(x, y, z, t)}{\partial z} \Big|_{z=0} = 0  \\[1ex]
T(x, y, z, 0) = T_\text{base} 
\end{dcases}
\label{HEq_1}
\end{equation}
In \eqref{HEq_1}, $\rho_m $ denotes the mass density, $C_p$ the specific heat capacity, 
and $k$ the heat conductivity of the skin. The quantity
$P_d^{(a)} \equiv (\alpha \cos\theta_1  P_d^{(i)})$ 
represents the power per unit area transmitted through the skin surface at the beam center. 
$\theta_1$ is the incident angle and $\theta_2$ the refracted angle of the electromagnetic beam. 
$\mu$ is the skin absorption coefficient at the relevant electromagnetic frequency.
\eqref{HEq_1} is the initial boundary 
value problem (IBVP) governing the evolution of the skin temperature $T(x, y, z, t)$. 
%

\section{Analytical Solution of Skin Temperature }
\subsection{Nondimensionalization}
We first establish appropriate scales to nondimensionalize the physical quantities in \eqref{HEq_1}. 
The electromagnetic heat source term in \eqref{HEq_1} exhibits exponential decay with depth. 
The characteristic decay depth, $z_s \equiv 1/\mu$, serves as the natural length scale in the depth direction. 
In the lateral directions, the heat source varies with $f(x, y; \sigma_\xi, \sigma_\eta, \phi_2)$, 
the 2D Gaussian relative distribution of power density projected onto the skin surface. 
It is tempting to select either the half major axis $\sigma_\xi$ or 
the half minor axis $\sigma_\eta$ in $f(x, y)$ 
as the length scale in the lateral directions. 
However, our objective is to develop a nondimensional formulation that enables studying
the effects of incident angle and beam spot geometry. To accomplish this,  the nondimensionalization process itself must be designed to be independent of both the incident angle and the specific geometry of the beam spot. This design requirement ensures that the influence of these parameters is fully captured in the resulting nondimensional formulation, allowing for meaningful comparative analysis.
To that end, we introduce a representative lateral length scale, $r_s$, which reflects 
the overall lateral variation but remains unchanged when the incident angle and/or the specific beam spot geometry are varies. 
Fixing $r_s$ allows for a consistent nondimensionalization process 
across a family of beams, preserving the effects of beam geometry variations 
in the system after nondimensionalization. 
For instance, in a class of beams with an intrinsic beam radius about $5\,\text{cm}$, we set $r_s = 5\,\text{cm}$. Even when the actual beam radius varies from $3\,\text{cm}$ to $7\,\text{cm}$ within this group, the lateral length scale remains fixed, and
the effects of these variations are fully captured in the resulting nondimensional systrem. 
For the same reason, we have set the depth scale to $1/\mu$, rather than
$(\cos\theta_2/\mu)$. 

For a 95 GHz electromagnetic beam, the absorption coefficient is 
$\mu = 6250\,\text{m}^{-1}$, yielding a characteristic depth scale of $z_s = 1/\mu = 0.16\,\text{mm}$
\cite{Walters_2000}. 
In most applications, the size of the beam spot cross-section is several centimeters or larger
\cite{Parker_2017, Parker_2017B, Parker_2021, Parker_2024}. 
Consequently, the lateral length scale is much larger than the depth scale: $r_s \gg z_s$. 
Since the effect of heat conduction is inversely proportional to 
the square of the length scale, its dominant effect of heat conduction occurs in the depth direction 
while its effect in the lateral directions is negligible. 
As a result, the time scale of heat conduction is determined by that in the depth direction:
$\displaystyle t_s = \frac{z_s^2}{(k/\rho_m C_p)} = \frac{\rho_m C_p}{k \mu^2}$.  

The temperature scale is defined as $\Delta T_s= (T_\text{act}-T_\text{base})$,  
where $T_\text{act}$ is the activation temperature of thermal nociceptors
\cite{Tillman_1995}. 
This scale represents the temperature increase required to activate nociceptors. 
Below, we list the scales for relevant physical quantities and the resulting nondimensional quantities. 
To facilitate the subsequent analysis, we use the simple notation $X$ without the subscript
$_\text{nondim}$ for all 
nondimensional quantities. For clarity, the original physical quantities are denoted by
$X_\text{phy}$. 
\begin{equation}
\begin{dcases}
 z_s \equiv \frac{1}{\mu}, \qquad z \equiv \frac{z_\text{phy}}{z_s} \\[0ex]
r_s \equiv (\text{lateral length scale}), \qquad 
(x, y) \equiv \frac{(x_\text{phy}, y_\text{phy})}{r_s} \\[0ex]
(\sigma_1, \sigma_2, \sigma_\xi, \sigma_\eta) \equiv 
\frac{1}{r_s}(\sigma_1, \sigma_2, \sigma_\xi, \sigma_\eta)_\text{phy} \\[1ex]
t_s \equiv \frac{\rho_m C_p}{k \mu^2},\qquad t \equiv \frac{t_\text{phy}}{t_s} \\[1ex]
T_s\equiv (T_\text{act,phy}-T_\text{base,phy}), \qquad 
T \equiv \frac{T_\text{phy}-T_\text{base,phy}}{T_s} \\[1ex]
P_s\equiv k \mu T_s, \quad P_\text{d}^{(i)} \equiv\frac{P_\text{d,phy}^{(i)}}{P_s}, 
\quad P_d^{(a)} \equiv\frac{P_\text{d,phy}^{(a)}}{P_s} = (\alpha \cos\theta_1  P_d^{(i)}) \\
f(x, y; \sigma_\xi, \sigma_\eta, \phi_2) \equiv f_\text{phy}(x_\text{phy}, y_\text{phy}; 
\sigma_{\xi,\text{phy}}, \sigma_{\eta,\text{phy}}, \phi_2) 
\end{dcases}
\label{scales_nd_qs} 
\end{equation}
Note that since all lateral length quantities are normalized by $r_s$, the functional  
form of $f(x, y)$ remains unchanged after nondimensionalization. 
Substituting \eqref{scales_nd_qs} into \eqref{HEq_1}, 
we obtain the nondimensional IBVP for $T(x,y,z,t)$. 
\begin{equation}
\begin{dcases}
\displaystyle \frac{\partial T}{\partial t} 
= \varepsilon^2 \Big(\frac{\partial^2T}{\partial x^2}
+\frac{\partial^2T}{\partial y^2}\Big)+\frac{\partial^2T}{\partial z^2}
+P_d^{(a)} {\cdot} f(x+\varepsilon z \tan\theta_2, y) 
\frac{1}{\lambda} e^{-z/\lambda}, 
\quad \lambda \equiv \cos\theta_2 \\[2ex]
\displaystyle \frac{\partial T(x, y, z, t)}{\partial z} \bigg|_{z=0}=0 \\[1ex]
\displaystyle T(x, y, z, 0) = 0 
\end{dcases}
\label{IBVP_1}
\end{equation}
where $\displaystyle \varepsilon \equiv \frac{z_s}{r_s} $ is the ratio of depth scale to lateral scale. 
In most applications, $\varepsilon \ll 1$. 
In the nondimensionalization process outlined in \eqref{scales_nd_qs}, all scalings are independent of 
the incident angle $\theta_1$ and independent of a particular beam geometry. 
As a result, the effects of $\theta_1$ and beam geometry are  
fully retained in the nondimensional IBVP presented in \eqref{IBVP_1}. 
Specifically, in \eqref{IBVP_1} the following items are affected by $\theta_1$. 
\begin{equation}
\begin{aligned}
& P_d^{(a)} = (\alpha \cos\theta_1  P_d^{(i)}), 
\quad \theta_2 = \arcsin(\frac{n_\text{air}}{n_\text{skin}}\sin\theta_1),
\quad \lambda = \cos\theta_2 \\
& f(x, y) = 
f\Big(x, y; \big[\sigma_\xi, \sigma_\eta, \phi_2 \big](\sigma_1, \sigma_2, \phi, \theta_1) \Big) 
\end{aligned}
\label{effect_q1}
\end{equation}

\subsection{Asymptotic Solution of \eqref{IBVP_1}}
The presence of the small parameter $\displaystyle \varepsilon \equiv \frac{z_s}{r_s} \ll 1$ in \eqref{IBVP_1} suggests a solution of the form 
\begin{equation}
T(x, y, z, t) = T^{(0)}(x, y, z, t) + \varepsilon T^{(1)}(x, y, z, t) + \cdots 
\label{asymp_form}
\end{equation}
where $T^{(0)}$ and $T^{(1)}$ represent the leading-order and first-order terms in the asymptotic expansion, respectively.
In this study, we focus on the leading-order term $T^{(0)}(x, y, z, t)$. 
Substituting the asymptotic solution \eqref{asymp_form} into the IBVP \eqref{IBVP_1} and balancing the $O(1)$ terms, 
we obtain the governing IBVP for $T^{(0)}(x,y,z,t)$:
\begin{equation}
\begin{dcases}
\displaystyle \frac{\partial T^{(0)}}{\partial t} 
= \frac{\partial^2 T^{(0)}}{\partial z^2}
+P_d^{(a)} f(x, y) \frac{1}{\lambda} e^{-z/\lambda }, \quad 
\lambda \equiv \cos\theta_2 \\[1ex]
\displaystyle \frac{\partial T^{(0)}(x, y, z, t)}{\partial z} \bigg|_{z=0}=0 \\[1ex]
\displaystyle T^{(0)}(x, y, z, 0) = 0 
\end{dcases}
\label{IBVP_2}
\end{equation}
The governing IBVP \eqref{IBVP_2} for $T^{(0)}(x,y,z,t)$ has several key properties: 
\begin{itemize}
\item The forcing term (the heat source) in the differential equation 
$P_d^{(a)} f(x, y) \frac{1}{\lambda} e^{-z/\lambda }$ has separate dependences on 
variable $z$ and variables $(x, y)$.
\item \eqref{IBVP_2} does not involve any derivatives with respect to $(x, y)$. 
As a result, \eqref{IBVP_2} can be viewed as an IBVP in variables $(z, t)$
with $(x, y)$ as parameters. 
\item The differential equation, boundary and initial conditions are linear in $T^{(0)}$.
\item The boundary and initial conditions are homogeneous. 
\end{itemize}
With the properties above, we can take care of the forcing term in \eqref{IBVP_2}
using the principle of superposition. It follows that the solution $T^{(0)}(x,y,z,t)$ of 
\eqref{IBVP_2} also has separate dependences on variable $z$ and variables $(x, y)$. 
Specifically, it has the form: 
\begin{equation}
 T^{(0)}(x, y, z, t) = P_d^{(a)} {\cdot} f(x, y) W^{(0)}(z, t; \lambda)
 \label{T0_P0W0}
\end{equation}
where $f(x,y)$ captures the spatial variation of skin temperature in lateral directions, and 
$W^{(0)}(z, t; \lambda)$ describes the evolution of skin temperature along the depth direction. 
$W^{(0)}$ is governed by the IBVP below in variables $(z, t)$ with 
$\lambda = \cos\theta_2$ as a parameter. 
\begin{equation}
\begin{dcases}
\displaystyle \frac{\partial W^{(0)}}{\partial t} 
= \frac{\partial^2 W^{(0)}}{\partial z^2}
+\frac{1}{\lambda} e^{-z/\lambda } \\[1ex]
\displaystyle \frac{\partial W^{(0)}(z, t)}{\partial z} \bigg|_{z=0}=0 \\[1ex]
\displaystyle W^{(0)}(z, 0) = 0 
\end{dcases}
\label{IBVP_3}
\end{equation}
In the scenario where the incident beam is perpendicular to the skin surface, we have
$\theta_1 = \theta_2 = 0$ and $\lambda = \cos\theta_2 =1$. 
Previously, we solved \eqref{IBVP_3} in the case of $\lambda=1$
\cite{WBZ_2020C}.
\begin{equation}
\boxed{\quad \begin{aligned}
& U^{(0)}(z, t) \equiv W^{(0)}(z, t; \lambda = 1) = -e^{-z}+\frac{e^{-z+t}}{2} 
\text{erfc}(\frac{-z+2t}{\sqrt{4t} }) \\
& \qquad \qquad + \frac{e^{z+t}}{2} \text{erfc}(\frac{z+2t}{\sqrt{4t} }) 
-z \, \text{erfc}(\frac{z}{\sqrt{4 t} }) 
+ \frac{2\sqrt{t}}{\sqrt{\pi }} e^{\frac{-z^2}{4t}}  
\end{aligned} \quad }
\label{U0_exp}  
\end{equation}
Note that $U^{(0)}(z, t)$ is a function of $(z, t)$ only, independent of any parameter. 
For the general case of $\lambda > 0$, we scale
variables $(z, t)$ and function $W^{(0)}(z, t; \lambda)$ as follows:
\begin{equation}
(\tilde{z}, \tilde{t}) \equiv (\frac{z}{\lambda}, \frac{t}{\lambda^2} ), \qquad 
\tilde{W}(\tilde{z}, \tilde{t}; \lambda) \equiv \frac{1}{\lambda} W^{(0)}(z, t; \lambda) 
\label{scaling_2}
\end{equation}
We use the chain rule to express derivatives of $W^{(0)}$ with respect to 
$(z, t)$ in terms of derivatives of $\tilde{W}$ with respect to $(\tilde{z}, \tilde{t})$. 
It is straightforward to verify that for any $\lambda > 0$, 
$\tilde{W}(\tilde{z}, \tilde{t}; \lambda) $ satisfies 
IBVP \eqref{IBVP_3} with $\lambda = 1$.
Thus, $\tilde{W}(\tilde{z}, \tilde{t}; \lambda) $ is independent of $\lambda $, 
and we have $\tilde{W}(\tilde{z}, \tilde{t}; \lambda) = U^{(0)}(\tilde{z}, \tilde{t})$. 
Using \eqref{scaling_2}, we express $W^{(0)}$ in terms of $U^{(0)}$: 
\begin{equation}
W^{(0)}(z, t; \lambda) = \lambda U^{(0)}(\tilde{z}, \tilde{t})
\Big|_{(\tilde{z}, \tilde{t})=(\frac{z}{\lambda}, \frac{t}{\lambda^2})}
\label{W0_U0} 
\end{equation}
Combining \eqref{T0_P0W0} and \eqref{W0_U0}, we obtain 
the leading-order asymptotic solution to the IBVP \eqref{IBVP_1}.
\begin{equation}
\boxed{\quad 
\begin{aligned}
& T^{(0)}(x, y, z, t; P_d^{(i)}, \sigma_1, \sigma_2, \phi, \theta_1 ) 
= P_d^{(a)} {\cdot} f(x, y; \sigma_{\xi}, \sigma_{\eta}, \phi_2) 
\, \lambda U^{(0)}(\tilde{z}, \tilde{t})
\Big|_{(\tilde{z}, \tilde{t})=(\frac{z}{\lambda}, \frac{t}{\lambda^2})} \\[1ex]
& \qquad P_d^{(a)} = (\alpha\, P_d^\text{(proj)}) = (\alpha \cos \theta_1 P_d^{(i)}), 
\qquad \lambda = \cos\theta_2 
\end{aligned} \quad} 
\label{T0_asymp}
\end{equation} 
In solution expression \eqref{T0_asymp}, $\theta_1$ is the incident angle of 
the electromagnetic beam with respect to the skin surface. The intrinsic beam parameters 
$(P_d^{(i)}, \sigma_1, \sigma_2, \phi)$ are independent of the incident angle. 
On the righthand side of \eqref{T0_asymp}, there are several derived parameters. 
\begin{itemize}
\item $P_d^{(a)}$: the beam center power density projected onto the skin surface and passing into the skin. $\alpha$ is the fraction of the arriving power passing into the skin. 
\item $f(x, y; \sigma_{\xi}, \sigma_{\eta}, \phi_2)$: the 2D Gaussian relative distribution of power density projected onto the skin surface as a function of normalized variables $(x, y)$, given in \eqref{Pd_eff}. 
\item $(\sigma_{\xi}, \sigma_{\eta}, \phi_2)$: the geometric parameters
describing the beam spot projected onto the skin surface. 
$(\sigma_{\xi}, \sigma_{\eta}, \phi_2)$ are calculated from the
intrinsic beam parameters $(\sigma_1, \sigma_2, \phi)$ and the incident angle $\theta_1$, 
as described in \eqref{phi2_s1_s2}. 
\end{itemize}
%

\section{Scaling Properties}
\subsection{Scaling in 3D Skin Temperature} \label{3D_temp_scaling}
We consider the two beam setups described in \eqref{projected_setup}: 
the original beam setup at a nontrivial incident angle
with parameters $(P_d^{(i)}, \sigma_1, \sigma_2, \phi, \theta_1)$, 
and the projected beam setup perpendicular to the skin surface with parameters 
$(P_d^\text{(proj)}, \sigma_\xi, \sigma_\eta, \phi_2, \theta_1= 0)$, where 
$P_d^\text{(proj)} = \cos\theta_1 P_d^{(i)}$ and 
$(\sigma_\xi, \sigma_\eta, \phi_2)$ are calculated from 
$(\sigma_1, \sigma_2, \phi, \theta_1)$ in \eqref{phi2_s1_s2}. 
The original and projected beams share the same power density
projected onto the skin surface.
They differ in incident angle, in intrinsic power density,  
and in intrinsic geometric parameters over a perpendicular beam cross-section. 

\begin{figure}[!h]
\vskip 0.6cm
\begin{center}
\psfig{figure=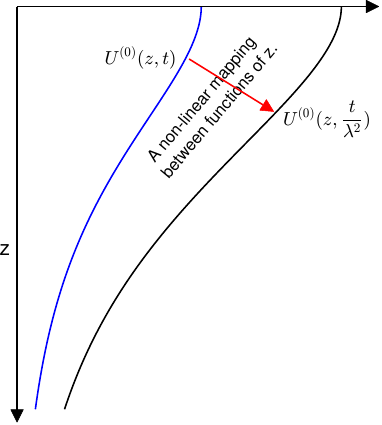, width=1.9in} \;
\psfig{figure=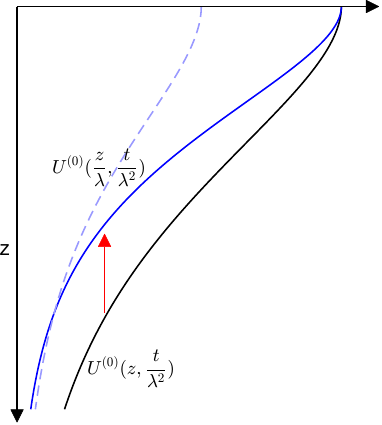, width=1.9in} \;
\psfig{figure=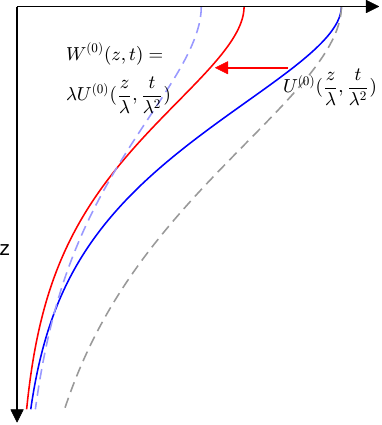, width=1.9in}
\end{center}
\vskip -0.6cm
\caption{Illustration of the mapping from $U^{(0)}(\bullet, t)$ to $W^{(0)}(\bullet, t)$
in \eqref{mapping_1}. 
Left: $U^{(0)}(\bullet, t) \rightarrow U^{(0)}(\bullet, \frac{t}{\lambda^2})$.
Center: $U^{(0)}(\bullet, \frac{t}{\lambda^2}) \rightarrow 
U^{(0)}(\frac{\bullet}{\lambda}, \frac{t}{\lambda^2})$. 
Right: $U^{(0)}(\frac{\bullet}{\lambda}, \frac{t}{\lambda^2}) 
\rightarrow W^{(0)}(\bullet, t)$.} 
\label{fig_04}
\end{figure}

For the original beam, the 3D skin temperature distribution 
as a function of $(x, y, z, t)$ is given directly in \eqref{T0_asymp}.
This solution is valid for arbitrary incident angle $\theta_1$ 
and arbitrary intrinsic beam parameters $(P_d^{(i)}, \sigma_1, \sigma_2, \phi)$. 
This solution expression is also applicable to the projected beam. 
For the projected beam, the incident angle is $\theta_1 = 0$, resulting in 
$\theta_2 = 0$ and $\cos\theta_2 = 1$. When this already projected beam 
is projected again onto the skin surface, both the power density and 
the beam spot geometry remain unchanged. 
\[ \begin{dcases}
(\alpha \cos\theta_1 P_d^\text{(proj)}) \Big|_{\theta_1 = 0} = (\alpha P_d^\text{(proj)}) \\
\big[\sigma_\xi, \sigma_\eta, \phi_2 \big](\sigma_\xi, \sigma_\eta, \phi_2, 0)
 = [\sigma_\xi, \sigma_\eta, \phi_2] 
\end{dcases} \]
It follows that for the projected beam, the 3D skin temperature distribution is 
\begin{equation}
T^{(0)}(x, y, z, t; P_d^\text{(proj)}, \sigma_\xi, \sigma_\eta, \phi_2, 0) 
= (\alpha P_d^\text{(proj)}) f(x, y; \sigma_\xi, \sigma_\eta, \phi_2) U^{(0)}(z, t) 
\label{T0_q_0}
\end{equation} 
The 3D temperature distribution \eqref{T0_asymp} for the original beam 
is analogous to \eqref{T0_q_0} for the projected beam. 
They are related through scalings in variables $(z, t)$ 
and in function value $U^{(0)}$. We compare them to highlight 
the similarities and the differences. 
\begin{equation}
\begin{dcases}
\overbrace{\; T^{(0,\text{orig})}(x, y, z, t)\; }^{\substack{\text{3D temperature for} \\
\text{\bf the original beam}}}
\equiv T^{(0)}(x, y, z, t; P_d^{(i)}, \sigma_1, \sigma_2, \phi, \theta_1 ) \\
\hspace*{3cm} = (\alpha P_d^\text{(proj)}) f(x, y; \sigma_\xi, \sigma_\eta, \phi_2) 
\; \lambda U^{(0)}(\tilde{z}, \tilde{t})
\Big|_{(\tilde{z}, \tilde{t})=(\frac{z}{\lambda}, \frac{t}{\lambda^2})} \\[1ex]
\overbrace{\;T^{(0,\text{proj})}(x, y, z, t)\; }^{\substack{\text{3D temperature for} \\
\text{\bf the projected beam}}}
\equiv T^{(0)}(x, y, z, t; P_d^\text{(proj)}, \sigma_\xi, \sigma_\eta, \phi_2, 0) \\
\hspace*{3cm} = (\alpha P_d^\text{(proj)}) f(x, y; \sigma_\xi, \sigma_\eta, \phi_2) 
\; U^{(0)}(z, t)  
\end{dcases} \label{compare_2Ts}
\end{equation}
When we treat them as functions of $(x, y, z, t)$, we can start from 
the temperature distribution of the projected beam 
$T^{(0,\text{proj})}(x, y, z, t)$ and obtain the temperature distribution of the original beam
$T^{(0,\text{orig})}(x, y, z, t)$ as follows. 
\begin{itemize}
\item Scale the $z$-direction by a factor of $\lambda \equiv \cos\theta_2$.
\item Scale the $t$-direction by a factor of $\lambda^2$. 
\item Scale the function value by a factor of $\lambda$. 
\end{itemize}
For example, a point $(\frac{z}{\lambda}, \frac{t}{\lambda^2})$ in 
$T^{(0,\text{proj})}(x, y, z, t)$ before scaling 
becomes the point $(z, t)$ after scaling. 
As functions of $(x, y, z, t)$, $T^{(0,\text{orig})}(x, y, z, t)$ and $T^{(0,\text{proj})}(x, y, z, t)$
are related by 
\begin{equation}
\boxed{\quad
T^{(0,\text{orig})}(x, y, z, t) = \lambda T^{(0,\text{proj})}(x, y, 
\frac{z}{\lambda}, \frac{t}{\lambda^2}), \qquad \lambda = \cos\theta_2 \quad }
\label{scaling_law_1}
\end{equation}
Scaling law \eqref{scaling_law_1} corresponds to the mapping from 
$U^{(0)}$ to $W^{(0)}$ in \eqref{W0_U0}. 
Note that $T^{(0,\text{proj})}$ and $U^{(0)}$ at time $\frac{t}{\lambda^2}$ 
are mapped to $T^{(0,\text{orig})}$ and $W^{(0)}$ at time $t$. 
When we compare $U^{(0)}$ and $W^{(0)}$ at the same time level $t$, 
we need to view \eqref{W0_U0} as a mapping between two functions of $z$, instead 
of a mapping between two functions of $(z, t)$. 
Specifically, we view it as a mapping from $U^{(0)}(\bullet, t)$ 
(a function of $z$ at time $t$) to $W^{(0)}(\bullet, t)$ (a function of $z$ at time $t$). 
We dissect the mapping from $U^{(0)}(\bullet, t)$ to $W^{(0)}(\bullet, t)$
into three steps below: 
\begin{equation}
\begin{aligned}
& U^{(0)}(\bullet, t) \xrightarrow{\;\; \text{This is a nonlinear mapping on a function of $z$.}\;\;} 
U^{(0)}(\bullet, \frac{t}{\lambda^2}) \\[1ex]
& U^{(0)}(\bullet, \frac{t}{\lambda^2}) 
\xrightarrow{\;\;\text{This is a linear scaling in variable $z$.}\;\;} 
U^{(0)}(\frac{\bullet}{\lambda}, \frac{t}{\lambda^2}) \\[1ex]
& U^{(0)}(\frac{\bullet}{\lambda}, \frac{t}{\lambda^2}) 
\xrightarrow{\;\;\text{This is a linear scaling in function value.}\;\;} 
\lambda U^{(0)}(\frac{\bullet}{\lambda}, \frac{t}{\lambda^2}) 
= W^{(0)}(\bullet, t)
\end{aligned} \label{mapping_1}
\end{equation}
The three steps of the mapping are illustrated in Figure \ref{fig_04}. 
Notice that the first step is nonlinear when viewed as a mapping 
between two functions of $z$. 

In summary,  as two functions of $(x, y, z, t) $, 
the temperature distribution of the original beam at incident angle $\theta_1$, 
$T^{(0,\text{orig})}(x, y, z, t) $, is obtained from the temperature distribution of the projected beam, $T^{(0,\text{proj})}(x, y, z, t)$, via scalings in variables $(z, t)$ and in function value $T$. 
When comparing the two distributions in space at a fixed time $t$, 
there is a nonlinear step mapping $T^{(0,\text{proj})}$ from time $t$ 
to time $\frac{t}{\lambda^2}$. 

\subsection{Scaling of Skin Surface Heating Rates}
Let $T^\text{(surf)}(t; P_d^{(i)}, \theta_1)$ denote the beam center temperature 
on the skin surface at time $t$, caused by a beam with intrinsic beam center 
power density $P_d^{(i)}$ at incident angle $\theta_1$. 
Notice that the beam center temperature is independent of $(\sigma_1, \sigma_2, \phi)$,  
the intrinsic geometric parameters of the beam spot. 
Intrinsic parameters $(\sigma_1, \sigma_2, \phi)$ affect the projected 
beam spot on the skin surface and the lateral distribution of 3D skin temperature, 
not the beam center temperature. 
Evaluating the 3D temperature solution \eqref{T0_asymp} 
at the origin $(0, 0, 0)$  and using the fact that $f(0, 0)=1$, we obtain the expression of 
$T^\text{(surf)}(t; P_d^{(i)}, \theta_1)$. 
\begin{equation}
\begin{aligned}
& T^\text{(surf)}(t; P_d^{(i)}, \theta_1) 
\equiv T^\text{(0)}(x, y, z, t; P_d^{(i)}, \theta_1) 
\Big|_{(x,y,z)=(0,0,0)} = P_d^{(a)}\lambda  h(\frac{t}{\lambda^2}) \\
& \qquad \qquad 
P_d^{(a)} \equiv (\alpha P_d^\text{(proj)}) = (\alpha \cos \theta_1 P_d^{(i)}), 
\qquad \lambda = \cos\theta_2 
\end{aligned} 
\label{T_surf_exp}
\end{equation}
where $h(t)$ is a parameter-free single-variable function defined as:   
\begin{equation}
h(t) \equiv U^{(0)}(z, t) \Big|_{z=0} = \text{erfc}(\sqrt{t}) e^t -1+\frac{2}{\sqrt{\pi}} \sqrt{t} 
\label{hs_def}
\end{equation}
\eqref{T_surf_exp} is the skin surface temperature for the original beam with 
power density $P_d^{(i)}$ and incident angle $\theta_1$. 
For the projected beam with power density $P_d^\text{(proj)}$ and 
incident angle $\theta_1=0 $, the skin surface temperature is 
\begin{equation}
T^\text{(surf)}(t; P_d^\text{(proj)}, 0) 
= \alpha P_d^\text{(proj)} h(t) \label{T_surf_exp2}
\end{equation}
Comparing \eqref{T_surf_exp} and \eqref{T_surf_exp2}, 
we obtain the scaling law for $T^\text{(surf)}$: 
\begin{equation}
\boxed{\quad T^\text{(surf)}(t; P_d^{(i)}, \theta_1)  
= \lambda T^\text{(surf)}(\frac{t}{\lambda^2}; P_d^\text{(proj)}, 0), 
\quad \lambda = \cos\theta_2 \quad }
\label{scaling_law_1B}
\end{equation}
Next we study the rates of skin surface temperature increase caused by the original beam and 
by the projected beam. 
For small $t$ and for large $t$, respectively, function $h(t)$ has 
the asymptotic behaviors below: 
\begin{equation}
\begin{dcases}
h(t) = t - \frac{4}{3\sqrt{\pi}} t^{3/2} + O(t^2)  \quad \text{for small $t$}\\
h(t) = \frac{2}{\sqrt{\pi}} \sqrt{t} -1 + O(\frac{1}{\sqrt{t}}) \quad \text{for large $t$}
\end{dcases}
\label{ht_asymp}
\end{equation}
In experiments, the skin surface temperature is recorded with an infrared thermal camera 
at a sequence of time instances \cite{Parker_2017, Parker_2017B}. 
We use the rate of measured skin surface temperature increase to estimate $P_d^{(a)}$, 
the power density projected onto the skin surface and passing into the skin
\cite{WBZ_2023}.  
\begin{align}
& \frac{d}{dt}T^\text{(surf)}(t; P_d^{(i)}, \theta_1)  = 
\frac{P_d^{(a)}}{ \cos\theta_2} \quad \text{for small $t$} 
\label{Pd_est_st} \\[1ex]
& (\sqrt{\pi t}) \frac{d}{dt}T^\text{(surf)}(t; P_d^{(i)}, \theta_1)  = 
P_d^{(a)} \quad \text{for large $t$} 
& \label{Pd_est_Lt}
\end{align}
If we use \eqref{Pd_est_Lt} for large $t$ to estimate $P_d^{(a)}$, 
the estimate is unaffected by the incident angle. 
In contrast, if we use \eqref{Pd_est_st} for small $t$ to estimate 
$P_d^{(a)}$ and we ignore the effect of incident angle,  
then $P_d^{(a)}$ is over-estimated by a factor of $(1/\cos\theta_2)$. 

\begin{figure}[!h]
\vskip 0.6cm
\begin{center}
\psfig{figure=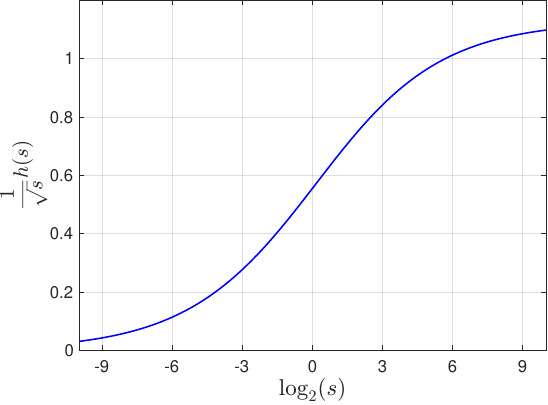, width=3in} \;\;
\psfig{figure=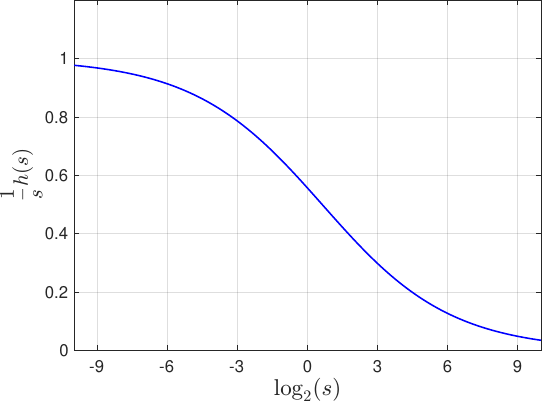, width=3in}
\end{center}
\vskip -0.6cm
\caption{Left: Graph of $\displaystyle \frac{1}{\sqrt{s}} h(s)$. 
Right: Graph of $\displaystyle \frac{1}{s} h(s)$.} 
\label{fig_05}
\end{figure}

We now present two mathematical assertions on function $h(t)$, which will be used 
to compare surface temperatures caused by several beam setups at the same time level $t$. 
\begin{itemize}
\setlength{\itemindent}{2em}
\item[(A1):] The function $\displaystyle \frac{1}{\sqrt{s}} h(s)$ is monotonically increasing for $s > 0$.
\item[(A2):] The function $\displaystyle \frac{1}{s} h(s)$ is monotonically decreasing for $s > 0$.
\end{itemize}
Assertions (A1) and (A2) can be derived analytically. Here, we demonstrate their validity
numerically. Figure \ref{fig_05} illustrates the graphs of $\displaystyle \frac{1}{\sqrt{s}} h(s)$
(left panel) and $\displaystyle \frac{1}{s} h(s)$ (right panel). 
The monotonic trend of each function is clearly evident in Figure \ref{fig_05}.

For $\lambda < 1$, we have $s=t/\lambda^2 > t$. Assertions (A1) and (A2) imply respectively 
\begin{align}
& \text{(A1):}\; \frac{1}{\sqrt{s}} h(s)\Big|_{s=t/\lambda^2} - \frac{1}{\sqrt{t}} h(t) > 0 
\quad \Longrightarrow \quad 
\frac{1}{\sqrt{t}} \Big( \lambda h(\frac{t}{\lambda^2}) - h(t) \Big) > 0
\nonumber \\
& \qquad \Longrightarrow \quad 
\boxed{\quad \lambda h(\frac{t}{\lambda^2}) > h(t) \quad \text{ for any $\lambda < 1$.} 
\quad} \label{h_res_1} \\[3ex]
& \text{(A2):}\; \frac{1}{t} h(t) -\frac{1}{s} h(s)\Big|_{s=t/\lambda^2} > 0 
\quad \Longrightarrow \quad 
\frac{1}{t} \Big(h(t)-\lambda^2 h(\frac{t}{\lambda^2}) \Big) > 0
\nonumber \\
& \qquad \Longrightarrow \quad 
\boxed{\quad h(t) > \lambda^2 h(\frac{t}{\lambda^2}) \quad \text{ for any $\lambda < 1$.} 
\quad} \label{h_res_2}
\end{align}
For $\lambda=\cos\theta_2$, utilizing expression \eqref{T_surf_exp} and scaling law \eqref{scaling_law_1B}, we interpret inequalities 
\eqref{h_res_1} and \eqref{h_res_2} in terms 
of the skin surface temperature $T^\text{(surf)}$ for three beam setups. 
\begin{align}
& T^\text{(surf)}(t; P_d^{(i)}, \theta_1) = 
\underbrace{\lambda T^\text{(surf)}(\frac{t}{\lambda^2}; P_d^\text{(proj)}, 0) 
> T^\text{(surf)}(t; P_d^\text{(proj)}, 0)}_{\lambda h(\frac{t}{\lambda^2}) > h(t)} 
\label{Tsurf_R1} \\[1ex]
& \underbrace{T^\text{(surf)}(t; P_d^{(i)}, 0) > \lambda^2 T^\text{(surf)}(\frac{t}{\lambda^2}; P_d^{(i)}, 0)}_{h(t) > \lambda^2 h(\frac{t}{\lambda^2})} 
= \lambda T^\text{(surf)}(\frac{t}{\lambda^2}; ( \lambda P_d^{(i)}), 0) \nonumber \\
& \qquad \qquad > \lambda T^\text{(surf)}(\frac{t}{\lambda^2}; P_d^\text{(proj)}, 0) 
= T^\text{(surf)}(t; P_d^{(i)}, \theta_1) 
\label{Tsurf_R2}
\end{align}
In \eqref{Tsurf_R2}, we have used 
\( \lambda P_d^{(i)} = \cos\theta_2 P_d^{(i)} > \cos\theta_1 P_d^{(i)} 
=  P_d^\text{(proj)} \).
Combining \eqref{Tsurf_R1} and \eqref{Tsurf_R2}, we compare the skin 
surface temperatures caused by three beam setups. 
\begin{equation}
\boxed{\quad \underbrace{T^\text{(surf)}(t; P_d^{(i)}, 0)}_{\substack{
\text{original beam}\\ \text{at 0 incident angle}}}
\; >\;  \underbrace{T^\text{(surf)}(t; P_d^{(i)}, \theta_1)}_{\substack{
\text{original beam}\\ \text{at incident angle $\theta_1$}}}
\; >\;  \underbrace{T^\text{(surf)}(t; P_d^\text{(proj)}, 0)}_{\substack{
\text{projected beam}\\ \text{at 0 incident angle}}} \quad }
\label{Tsurf_res}
\end{equation}
In \eqref{Tsurf_res}, $P_d^{(i)}$ is the intrinsic beam center power density over 
a perpendicular beam cross-section of the original beam. 
$P_d^\text{(proj)} = \cos\theta_1 P_d^{(i)}$ is the beam center power density 
of the original beam projected on the skin surface,  which is also the 
beam center power density of the projected beam. 
\eqref{Tsurf_res} compares the skin surface temperatures at beam center at time $t$ 
caused by three beam setups. 
The term in the middle is the surface temperature for the original beam setup 
(with intrinsic power density $P_d^{(i)}$ at incident angle $\theta_1$).
The term on the right is the surface temperature for the projected beam setup 
(with power density $P_d^\text{(proj)} = \cos\theta_1 P_d^{(i)}$ at zero incident angle).
The term on the left is the surface temperature caused by the beam 
with power density $P_d^{(i)}$ at zero incident angle. This is the same intrinsic beam 
as the original beam but at zero incident angle. 
Inequality \eqref{Tsurf_res} states that, at any fixed time $t$, 
the surface temperature caused by a beam at incident angle $\theta_1$ 
is lower than the surface temperature caused by the same intrinsic beam at zero incident angle,
but is higher than the surface temperature caused by the projected beam at zero incident angle. 

\subsection{Scaling of Activated Skin Volume}
Thermal nociceptors in the skin are activated whenever the local skin temperature 
exceeds the activation threshold, denoted as $T_\text{act,phy}$. 
This activation threshold is used in the definition of temperature scale
$T_s \equiv T_\text{act,phy}-T_\text{base,phy}$ in \eqref{scales_nd_qs}.
After nondimensionalization, the nondimensional activation temperature 
becomes $T_\text{act}=1$. 
Let $V_\text{act}(t)$ denote the volume of the skin region where thermal nociceptors are activated at time $t$. 
\begin{equation}
 V_\text{act}(t) \equiv \text{volume}\Big\{(x, y, z) \Big| 
T^{(0)}(x, y, z, t) \ge 1 \Big\} 
\label{vact_def}
\end{equation}
We consider two beam setups described in \eqref{projected_setup}: 
the original beam setup with parameters $(P_d^{(i)}, \sigma_1, \sigma_2, \phi, \theta_1)$ 
and the projected beam setup with parameters 
$(P_d^\text{(proj)}, \sigma_\xi, \sigma_\eta, \phi_2, \theta_1= 0)$. 
As functions of $(x, y, z, t)$, the temperature distributions of the two beam setups 
are related by the scalings in \eqref{scaling_law_1}: 
$T^{(0,\text{orig})}(x, y, z, t) = \lambda T^{(0,\text{proj})}(x, y, 
\frac{z}{\lambda}, \frac{t}{\lambda^2})$. 

In scaling law \eqref{scaling_law_1}, the scaling in $z$ translates to a scaling in the activated volume. 
The scaling in $t$ relates the activated volumes at two different times.
The activated skin volume is defined by the condition $T^{(0)}(x, y, z, t) \ge 1$ 
in \eqref{vact_def}, which is clearly not linear in $T$. 
As a result, the scaling in $T$ does not produce a simple relation on the activated volume
in a straightforward way. 
Notice that as given in \eqref{T0_q_0}, 
the skin temperature for the projected beam is proportional to $P_d^\text{(proj)}$. 
To connect the activated skin volumes caused by the two beam setups, 
in \eqref{scaling_law_1} we move the scaling multiplier $\lambda = \cos\theta_2$ 
on temperature $T$ into the solution expression as a scaling multiplier on power density
$P_d^\text{(proj)}$. 
We consider the projected beam with scaled power density, described by parameters 
\(\big( \lambda P_d^\text{(proj)}, \sigma_\xi, \sigma_\eta, \phi_2, \theta_1= 0 \big)\). 
 The temperature distributions of the original beam and the projected beam with 
 scaled power density are related by scalings in $(z, t)$ only, with no scaling in $T$. 
 Specifically, we have:
\begin{equation}
T^{(0)}(x, y, z, t; P_d^{(i)}, \sigma_1, \sigma_2, \phi, \theta_1 ) 
= T^{(0)}(x, y, \frac{z}{\lambda}, \frac{t}{\lambda^2}; 
\lambda P_d^\text{(proj)}, \sigma_\xi, \sigma_\eta, \phi_2, 0) 
\label{scaling_law_1M}
\end{equation}
Scaling law \eqref{scaling_law_1M} leads to a scaling law on the activated volumes 
of the original beam and the projected beam with scaled power density as two functions of $t$. 
\begin{equation}
\begin{aligned}
& V_\text{act}(t; P_d^{(i)}, (\sigma_1\sigma_2), \theta_1 ) 
= \lambda V_\text{act}(\frac{t}{\lambda^2}; \; \lambda P_d^\text{(proj)}, 
(\sigma_\xi \sigma_\eta), 0) 
\end{aligned}
\label{scaling_law_2}
\end{equation}
Note that, among the beam's geometric parameters, 
the activated volume depends only on $(\sigma_\xi \sigma_\eta)$, 
the area of the beam spot projected onto the skin surface. 
The azimuthal orientation angle $\phi_2$ rotates the activated skin region but does not 
alter its volume.
When the beam spot area is kept fixed, 
changing the beam aspect ratio $\sigma_\xi/\sigma_\eta$ stretches the activated skin region 
in one direction and shrinks it simultaneously by the same factor in the perpendicular direction, 
leaving the volume unchanged. 
The area of the projected beam spot on the skin surface is 
$\sigma_\xi \sigma_\eta = \sigma_1 \sigma_2/\cos\theta_1 $. 
Substituting $\lambda = \cos\theta_2$, $P_d^\text{(proj)} =\cos\theta_1 P_d^{(i)}$
and $\sigma_\xi \sigma_\eta = \sigma_1 \sigma_2/\cos\theta_1 $ 
into \eqref{scaling_law_2}, we obtain the scaling law below. 
\begin{equation}
\boxed{\quad
\underbrace{V_\text{act}(t; P_d^{(i)}, (\sigma_1\sigma_2), \theta_1 )}_{\substack{
\text{original beam at incident} \\ \text{angle $\theta_1$ at time $t$}}}
= \underbrace{V_\text{act}\Big(\frac{t}{\cos^2\theta_2}; \; 
(\cos\theta_2\cos\theta_1 P_d^{(i)}), (\frac{\cos\theta_2}{\cos\theta_1} \sigma_1 \sigma_2), 
0 \Big) }_{\substack{
\text{modified beam at zero incident} \\ \text{angle at time $(t/\cos^2\theta_2)$}}}
 \quad }
\label{scaling_law_2B}
\end{equation}
For a beam at incident angle $\theta_1$, scaling law \eqref{scaling_law_2B}
states that its activated volume at time $t$ is equal to the activated volume of a modified beam 
at an extended exposure time $( t/\cos^2\theta_2)$. 
In the modified beam, the incident angle is zero. The
intrinsic beam center power density is scaled by a factor of 
$\cos\theta_2\cos\theta_1 < 1$, and the intrinsic beam spot area is 
scaled by a factor of $\cos\theta_2/\cos\theta_1 > 1$. Here 
$\theta_2 $ is the refracted angle corresponding to the incident angle $\theta_1$. 

We examine the time at which a given threshold on the activated skin volume is reached. 
As noted earlier, the activated volume is proportional to the beam spot area. 
In the case of very large beam spot area, the time to reach the volume threshold 
is approximately the time when the activation temperature $T_\text{act}$ is reached. 
In the temperature distribution \eqref{T0_asymp}, the spatial maximum occurs 
at the beam center on the skin surface, which is denoted by 
$T^\text{(surf)}$ and is given in \eqref{T_surf_exp}. 
Before $T^\text{(surf)}$ reaches $T_\text{act}$, no skin region is activated. 
As time $t$ increases, $T^\text{(surf)}$ increases monotonically. 
When $T^\text{(surf)}$ goes slightly above $T_\text{act}$, 
the skin within a small depth becomes activated. 
This small activated depth, when multiplied by the very large beam spot area, produces 
an activated volume above the volume threshold.
Let $t_\text{act}$ denote the activation time, defined as the time when $T^\text{(surf)}$
reaches $T_\text{act}$ and the activated volume transitions from zero to positive. 
\begin{equation}
t_\text{act}(P_d^{(i)}, \theta_1) \equiv 
\min \Big\{ t \; \Big| T^\text{(surf)}(t; P_d^{(i)}, \theta_1)  \ge 
T_\text{act} \Big\} 
\label{t_pv_def}
\end{equation}
In the case of very large beam spot area, $t_\text{act}(P_d^{(i)}, \theta_1)$ 
is approximately the time when the activated volume reaches the volume threshold.
In \eqref{Tsurf_res}, we compared the skin surface temperatures at time $t$ caused by three 
beam setups. The ordering of skin surface temperatures at fixed time $t$ 
given in \eqref{Tsurf_res} leads to the corresponding ordering of activation times 
for the three beam setups. 
\begin{equation}
\boxed{\quad \underbrace{t_\text{act}(P_d^{(i)}, 0) }_{\substack{
\text{original beam}\\ \text{at 0 incident angle}}}
\; <\;  \underbrace{t_\text{act}(P_d^{(i)}, \theta_1)}_{\substack{
\text{original beam}\\ \text{at incident angle $\theta_1$}}}
\; <\;  \underbrace{t_\text{act}(P_d^\text{(proj)}, 0)}_{\substack{
\text{projected beam}\\ \text{at 0 incident angle}}} \quad }
\label{t_pv_res}
\end{equation}

In \eqref{t_pv_res}, the term in the middle is the activation time 
for the original beam setup 
(with intrinsic power density $P_d^{(i)}$ at incident angle $\theta_1$).
The term on the right is the activation time for the projected beam setup 
(with power density $P_d^\text{(proj)} = \cos\theta_1 P_d^{(i)}$ at zero incident angle).
The term on the left is the activation time for the beam 
with power density $P_d^{(i)}$ at zero incident angle. This is the same intrinsic beam 
as the original beam but at zero incident angle. 
Inequality \eqref{t_pv_res} states that for a given beam at incident angle $\theta_1$, 
the time required to reach thermal nociceptor activation is longer than the activation time 
for the same beam at zero incident angle but is shorter than the activation time 
for the projected beam at zero incident angle. 
%

\section{Conclusions}
We studied the thermal effect on skin from an electromagnetic beam at an incident angle. 
We carried out nondimensionalization and solved the nondimensional formulation 
for a leading-term asymptotic solution. The asymptotic analysis is not based on 
that the incident angle is small. Rather it is based that the depth scale is much smaller 
than the lateral scale. 
For a millimeter wavelength beam, the penetration depth of the electromagnetic wave is 
sub-millimeter, leading to concentrated heating in the top skin layer and
a large temperature gradient in the depth direction. 
In contrast, in the lateral directions, the characteristic length is described by the size of beam spot, which spans at least several centimeters in applications. This separation of length scales 
in the depth and lateral directions makes the ratio of depth to lateral length scales 
a small parameter in the nondimensional formulation. 
One consequence of this small parameter is that the heat conduction in the lateral 
directions is negligible compared to that in the depth direction.

When skin is exposed to an electromagnetic beam at an incident angle, the beam's power is first projected onto the skin surface. This projection affects the power per area over the skin surface, 
making it different from the power density over a perpendicular beam cross-section. 
This projection is included in the thermal model. 
A fraction of the power arriving at the skin surface is absorbed into the skin 
and continues propagating inside the skin along the refracted angle. 
Due to the refracted angle, the beam propagation distance is more than the depth 
and the power density absorption per depth is more than that in the case of zero refracted angle. 
Also the power density (and thus the heating source) shifts 
laterally as the beam propagates into the skin. 
This lateral shift is comparable to the depth scale but much smaller than the lateral length scale.
The asymptotic analysis is based on the ratio of the depth scale to the 
lateral length scale being small, not on the incident angle being small. 
The asymptotic solution obtained is expressed in terms of parameter-free 
single-variable functions and is valid for any incident angle. 

Based on the asymptotic solution, we investigated the thermal effects of three beam setups. Each beam setup is defined by the intrinsic parameters (power density, beam spot geometry)
over a perpendicular beam cross-section and the incident angle. 
We examined the following three beam setups:
\begin{itemize}
\item Beam 1: the original beam at incident angle $\theta_1$.
\item  Beam 1v: the beam with the same intrinsic parameters as the original beam 
but at zero incident angle (i.e., the original intrinsic beam at zero incident angle).
\item Beam 2: the projected beam at zero incident angle. The projected beam has the same power density projected onto the skin surface as the original beam (beam 1). 
\end{itemize}
As illustrated in Figure \ref{fig_02B}, beams 1 and 2 
share the same power density projected on the skin surface but 
have different incident angles. 
beams 1 and 1v share the same intrinsic power density over their respective 
perpendicular cross-section but have different incident angles. 
beam 2 has a lower power density and a larger beam spot area than beam 1v 
although both are at zero incident angle. 
From the analysis, we drew the following conclusions. 
\begin{enumerate}
\item
For the original beam at incident angle $\theta_1$ (beam 1), the electromagnetic heating source 
inside the skin decays in the depth direction faster than beams 1v and 2, which are both 
perpendicular to the skin surface. This is because the propagation distance per depth is longer 
for a beam ray along a refracted angle than for one along the depth directly. 
This increased path length results in greater absorption and attenuation of 
the electromagnetic wave as it propagates into the skin.
\item The 3D skin temperature distribution as a function of $(x, y, z, t)$ caused by 
the original beam (beam 1) is related to that caused by the projected beam (beam 2)
through scalings in variables $(z, t)$ and in temperature $T$. 
In particular, the temperature distribution of beam 1 at time $t$ is related to 
that of beam 2 at the extended exposure time $(t/\cos^2\theta_2)$, 
where $\theta_2$ is the refracted angle corresponding to the incident angle $\theta_1$.  
\item 
In exposure tests, the skin surface temperature is recorded at discrete time instances using 
an infrared thermal camera . 
For a beam perpendicular to the skin surface, the absorbed power density 
can be estimated from the measured rate of surface temperature increase 
at small times ($t$). 
In millimeter-wave (MMW) exposures, small $t$ refers to the time period
immediately after the start of exposure, within a fraction of the time scale. 
The time scale is the characteristic time of heat diffusion over the depth scale.
Large $t$ refers to time durations several multiples of this time scale after the 
start of exposure. 
For a beam at an incident angle, estimating the absorbed power density 
from the surface temperature increase rate at small $t$ leads to an overestimate unless the incident angle is known and explicitly accounted for in the calculation.
In contrast, using the surface temperature increase rate at large $t$ yields an 
accurate estimate of the absorbed power density regardless of the incident angle. 
\item
We analytically compared the skin surface temperatures induced by the three beam setups.
At any fixed time $t$, the beam center temperature for the original beam (beam 1) is higher 
than the temperature for the projected beam (beam 2) but is lower than 
the temperature for beam 1v (the original intrinsic beam at zero incident angle). 
Given a beam with fixed intrinsic parameters, the most effective 
way to increase skin surface temperature is to set the incident angle to zero during exposure. 
On the other hand, when the power density projected onto the 
skin surface is fixed, which can be achieved by various beam setups, 
the skin surface temperature is higher if this fixed projected power density is 
achieved by a beam at a larger incident angle. 
\item 
In MMW exposures, thermal nociceptors in the skin are activated wherever the local 
temperature exceeds the activation threshold. 
The activated skin volume may be used to quantify the heat sensation transduced 
in the brain from nociceptive signals. 
The activated skin volume caused by the original beam at incident angle $\theta_1$ 
at time $t$ is equal to that caused by a modified beam 
at zero incident angle at an extended exposure time of $t/\cos^2\theta_2$, 
where $\theta_2$ is the refracted angle. 
In the modified beam, the intrinsic power density is reduced (multiplied by a 
factor of $\cos\theta_2 \cos\theta_1<1$), and the intrinsic beam spot area is slightly increased (multiplied by a factor of $\cos\theta_2/ \cos\theta_1>1$).
We also examined the activation time, defined as the time when the thermal nociceptor 
activation temperature is first reached. For the original beam at incident angle $\theta_1$
(beam 1), the activation time is longer than that for the same intrinsic beam 
at zero incident angle (beam 1v), 
but is shorter than that for the projected beam (beam 2).
\end{enumerate}
%

\clearpage
\noindent{\bf \large Acknowledgement and disclaimer}

\noindent \indent
The authors acknowledge the Joint Intermediate Force Capabilities Office of U.S. Department of Defense and the Naval Postgraduate School for supporting this work. The views expressed in this document are those of the authors and do not reflect the official policy or position of the Department of Defense or the U.S. Government.


\end{document}